\documentclass{article}

\usepackage{amsmath}
\usepackage{comment}
\usepackage{subcaption}
\usepackage{todonotes}
\usepackage{array}
\usepackage{booktabs}
\usepackage{url}
\usepackage{hyperref}

\usepackage[preprint]{neurips_2026}

\usepackage[utf8]{inputenc} % allow utf-8 input
\usepackage[T1]{fontenc}    % use 8-bit T1 fonts
\usepackage{hyperref}       % hyperlinks
\usepackage{url}            % simple URL typesetting
\usepackage{booktabs}       % professional-quality tables
\usepackage{amsfonts}       % blackboard math symbols
\usepackage{nicefrac}       % compact symbols for 1/2, etc.
\usepackage{microtype}      % microtypography
\usepackage{xcolor}         % colors
\usepackage{enumitem}
\usepackage{graphicx}
\usepackage{multirow}
\usepackage{tabularx}

\RequirePackage{listings}
\definecolor{dkgreen}{rgb}{0,0.6,0}
\definecolor{gray}{rgb}{0.5,0.5,0.5}
\definecolor{mauve}{rgb}{0.58,0,0.82}

\usepackage{hyperref}
\usepackage{doi} % Automatically turns doi={...} fields into hyperlinks



\title{ChemReporter: A Framework for Curating and Exporting Large-Scale Chemical Datasets for MLIP Training}

\author{%
  Marie Bluntzer \And
  Jules Tilly \And
  Christoph Brunken \AND
  {\normalfont InstaDeep} \\
  1 Triton Square, London, NW1 3BF, United Kingdom \\
  \texttt{\{m.bluntzer,c.brunken\}@instadeep.com} \\
}

\begin{document}

\maketitle

\begin{abstract}
Training set quality and diversity are key determinants of the reliability of machine learning interatomic potentials (MLIPs), yet using massive datasets in full is often impractical and redundant, making intelligent data selection essential. A major bottleneck, however, is the lack of infrastructure for uniformly accessing, curating, and subsampling heterogeneous large-scale chemical datasets, which differ widely in structure, metadata, and file format.
We address this gap with \textit{ChemReporter}, a modular, method-agnostic framework that converts arbitrary molecular and materials datasets into a unified, queryable representation and exports the results directly into MLIP-ready training data. \textit{ChemReporter} operates in three decoupled stages: \emph{processing}, which parses raw datasets into a partitioned Apache Parquet repository enriched with structural, physical, and chemical metadata; \emph{querying}, which filters and samples this repository via a CLI or Python API using arbitrary selection criteria, from simple physical constraints to custom, user-defined strategies; and \emph{exporting}, which streams the selected subset into an HDF5 file ready for direct use in modern MLIP training frameworks. Throughout this process, every exported data point remains traceable to its original source entry, and dataset exports can be reliably reproduced given the same configuration and query database version. Because data is stored in a queryable, disk-backed format, \textit{ChemReporter} can process datasets far larger than available memory, allowing it to scale to billion-structure datasets on standard compute infrastructure.
\textit{ChemReporter} is available on \href{https://github.com/instadeepai/chemReporter}{GitHub} and \href{https://pypi.org/project/chemreporter}{PyPI} under the Apache License 2.0.
\end{abstract}

\section{Introduction}

Machine learning interatomic potentials (MLIPs) are a rapidly growing area of research~\cite{Bruntonetal2025}. They serve as a foundational tool for advanced molecular modeling in chemistry and materials science, bridging the accuracy of electronic structure calculations with the scalability of classical force fields~\cite{Jacobsetal2025}. To meet the demands of these complex tasks, recent software developments have focused on building highly optimized, scalable open-source libraries that accelerate model inference and distributed training~\cite{brunken2026machine}. These architectures are now widely used to study materials and biological systems~\cite{ChoJeon2026}. Yet effectively capturing diverse chemical spaces remains difficult, since MLIPs still struggle to generalize to chemical environments that differ from those seen during training~\cite{KreimanKrishnapriyan2025}.

The growing availability of large molecular and materials datasets, ranging from curated quantum chemical datasets to extensive molecular dynamics (MD) trajectories, has substantially advanced the field. With the emergence of massive collections such as the Open Molecules 2025 (OMol25)~\cite{levine2025openmolecules2025omol25} and Open Catalyst 2020 (OC20)~\cite{chanussot2021open} datasets, however, the community recognizes that poor quality data, and badly constructed datasets, can significantly harm the performance and generalization of models. Even in carefully constructed, widely used datasets, structural and numerical artifacts have been shown to persist and to measurably degrade downstream model behavior: for instance, data quality issues traced back to the underlying electronic structure calculations in SPICE were found to bias predicted liquid water density in MACE-OFF models~\cite{kovacs2025maceofftransferableshortrange}. Improving the reliability of MLIPs increasingly depends on the ability to efficiently construct, curate, and extract information-rich subsets from these vast data resources~\cite{zeng2026developing,Batzner2022}.

In this context, developing computational tools to manage, curate, filter, and actively select large-scale chemical datasets is essential. Such tools enable efficient exploration of high-dimensional structural space, reduce redundancy in quantum-mechanical labeling, and ensure coverage of physically relevant configurational landscapes. Ultimately, the reliability and generalization of MLIPs does not only depend on the raw size of training datasets but also on how effectively these datasets are constructed to represent the full diversity of the relevant chemical and configurational space. We note, however, that such curation must never be used to artificially inflate apparent model performance, but only to remove genuinely poor-quality data or to deliberately constrain a model's training distribution in a documented, traceable way (see Section~\ref{sec:curation} for further discussion).

To address this need, we introduce \textit{ChemReporter}, a modular software framework for pre-processing large-scale chemical datasets into a structured, queryable tabular format. \textit{ChemReporter} converts raw datasets into columnar tables stored in the Apache Parquet format, enriching each data point with a broad set of precomputed structural, compositional, and physical properties. This representation is extensible, allowing new properties or descriptors to be added as needed, and provides a unified basis from which users can query, filter, and subsample data according to their own selection criteria.
\textit{ChemReporter} already ships with a set of built-in filtering and selection methods, including checks for common structural artifacts such as implausible bonding, disconnected or floating atoms, and unphysically short interatomic distances, and this collection is intended to grow over time; however, its core design goal is to expose the underlying data as an easily queryable table, so that users can just as readily implement and apply their own custom selection strategies.

Beyond selection flexibility, \textit{ChemReporter} is designed to support reproducibility and traceability throughout data preparation. Every data point in an exported dataset retains a primary key linking it back to its original entry in the source dataset (e.g., OC20 or OMol25), allowing structures in a curated training set to be traced to their provenance. Likewise, because queries, sampling parameters, and random seeds are fully specified in a configuration file, a dataset export can be reliably reconstructed later, provided the same configuration and query database version are used and preserved by the user. Together, these properties make dataset construction with \textit{ChemReporter} auditable and repeatable, addressing a practical concern that is often overlooked in ad hoc data curation workflows.

Moreover, by leveraging Parquet's columnar storage, the framework enables fast, memory-efficient data manipulation: even billion-scale datasets can be queried, filtered, and subsampled without requiring full in-memory loading. Structurally, \textit{ChemReporter} acts as a method-agnostic data management layer that seamlessly couples massive datasets like FairChemDB and Open Catalyst with any downstream selection paradigm, including diversity-based pruning, descriptor-space clustering, or physics-informed filtering. By standardizing these operations and offering native integration with cloud storage backends like AWS S3, \textit{ChemReporter} substantially accelerates data preparation, making large-scale, high-fidelity MLIP data workflows practical and accessible on standard compute infrastructure.

\section{Related Work}
\subsection{Existing Datasets for MLIP Development}

The development of MLIPs has historically been shaped by the availability of specialized quantum-mechanical datasets. Early efforts focused on gas-phase molecular datasets such as QM9~\cite{ramakrishnan2014quantum}, and later ANI-1x/ANI-1ccx~\cite{smith2018less}, which introduced large-scale conformational diversity but remained restricted to a narrow set of light elements. Subsequent datasets broadened this scope: AIMNet2~\cite{zubatyuk2024aimnet2} extended elemental coverage across the periodic table, Transition1x~\cite{schreiner2022transition1x} targeted reactive pathways involving transition metals, and SPICE/SPICE2~\cite{Eastman2023, Eastman2024} expanded coverage to non-covalent interactions and solvated biomolecular systems at high levels of theory.

In parallel, materials-focused datasets such as the Materials Project Trajectory (MPTrj) dataset~\cite{deng2023materials}, OQMD~\cite{kirklin2015open}, and AFLOW~\cite{taylor2014restful} provided millions of bulk material relaxation trajectories under periodic boundary conditions, primarily to support crystalline stability and property prediction. Complementary efforts captured non-equilibrium and dynamical behavior: MD17/rMD17~\cite{christensen2020role} provided ab-initio molecular dynamics trajectories for small molecules, while the large-scale sAlex dataset~\cite{batzner2024omat24} extended this coverage to heterogeneous, non-equilibrium configurations to mitigate distributional drift in long-timescale simulations.

While these datasets established the foundation for atomistic machine learning, they were computed under inconsistent quantum-chemical settings (e.g., differing functionals, pseudopotentials, and convergence criteria), limiting cross-dataset compatibility. To address this, Meta FAIR introduced five large-scale, unified datasets -- \textit{OMol25}, \textit{OC20-22}, \textit{OMat24}, \textit{ODAC23}, and \textit{OMC25} -- collectively spanning over half a billion 3D structures and forming the basis of foundation models such as UMA~\cite{wood2026umafamilyuniversalmodels}. \textbf{OMol25}~\cite{levine2025openmolecules2025omol25} unifies organic molecular chemistry -- including isolated molecules, solvated systems, reactive pathways, and biomolecular structures -- at a consistent, high level of theory ($\omega\text{B97M-V/def2-TZVPD}$), incorporating and extending prior datasets such as SPICE, QM9, Transition1X, RGD1 and ANI-1x. \textbf{OC20}~\cite{chanussot2021open} and \textbf{OC22}~\cite{Tran_2023} focus on catalytic surface chemistry, providing relaxations of molecules adsorbed on crystalline catalyst surfaces. \textbf{OMat24}~\cite{batzner2024omat24} extends bulk materials modeling with single-point calculations, ab-initio molecular dynamics trajectories, and structural relaxations, building on datasets such as Alexandria~\cite{cavignac2026aidrivenexpansionapplicationalexandria}, MPTrj~\cite{deng2023materials}, and the Materials Project~\cite{Jain2013}. \textbf{ODAC23}~\cite{shuaibi2023opendac23} targets metal-organic frameworks for carbon dioxide adsorption. \textbf{OMC25}~\cite{gharakhanyan2026open} captures the conformational and packing landscape of molecular crystals, including experimentally derived structures from the Cambridge Structural Database (CSD).

Beyond individual datasets, community-wide efforts such as NOMAD~\cite{Scheidgen2023}, JARVIS~\cite{choudhary2025jarvis}, the Materials Project~\cite{Jain2013}, ColabFit~\cite{fuemmeler2024advancing, tadmor2023colabfit}, and OPTIMADE~\cite{evans2024developments} have sought to aggregate, standardize, and provide unified access to this heterogeneous data landscape, underscoring the field's growing need for infrastructure that can handle chemical data at scale and across sources.

\subsection{Chemical Descriptors}

The curation, filtering, and sampling of large atomistic datasets rely on chemical descriptors that summarize structural and chemical information efficiently and without redundancy. These are broadly divided into 2D topological invariants and 3D geometric representations, each serving distinct roles in MLIP training and downstream molecular discovery. For dataset selection specifically, 3D descriptors are generally required to be invariant and equivariant with respect to the relevant physical symmetries (translation, rotation, and permutation). Beyond these two classes, QSAR-derived properties -- such as solubility, lipophilicity (logP), or polar surface area (TPSA) -- are widely used in drug discovery to predict macroscale molecular properties, but remain rarely applied to dataset selection for MLIP training, despite offering a complementary, application-relevant signal for characterizing dataset coverage.

\subsubsection{2D Topological Invariants and Virtual Screening}

2D topological fingerprints are fast to compute and are widely used for structural down-selection and virtual screening. The most common approach relies on Extended-Connectivity Fingerprints (ECFPs), derived from the Morgan algorithm and formalized by Rogers and Hahn~\cite{rogers2010extended}. ECFPs represent molecules as graphs, assigning atoms initial identifiers based on periodic properties, then iteratively updating them by aggregating neighboring-shell information up to a fixed bond radius (typically $r=2$ or $r=3$), and hashing the result into a fixed-length bit vector (e.g., 1024 or 2048 bits). Because ECFPs discard 3D coordinates, they are invariant to conformer noise, making them well suited for indexing large numbers of ground-state configurations, computing bitwise similarity (e.g., Tanimoto), and pruning redundant molecular topologies prior to more expensive geometric computations.

\subsubsection{3D Geometric and Local Environment Descriptors}

Hand-crafted 3D descriptors, such as SOAP~\cite{bartok2013representing} and ACE~\cite{drautz2019atomic}, encode local atomic environments using symmetry-invariant expansions of neighbor density (spherical harmonics and radial functions for SOAP, many-body permutation-invariant polynomials for ACE), providing a rigorous, continuous measure of structural similarity and diversity. These descriptors are widely used to identify under-sampled regions of configurational space and to guide unsupervised clustering or sampling strategies. More recently, learned representations from equivariant graph neural networks, such as SchNet~\cite{schutt2018schnet}, NequIP~\cite{Batzner2022}, and MACE~\cite{batatia2023macehigherorderequivariant}, provide an alternative, data-driven route to capturing geometric and directional structure, and are increasingly used in active learning pipelines, where prediction variance across model ensembles serves as an uncertainty signal for selecting new configurations for refinement.

% THIS can only be added when ACE/SOAP will be in chemreporter ... : To the best of our knowledge, no existing framework allows users to jointly query datasets using 3D physical descriptors, 2D topological fingerprints, and arbitrary custom-defined properties within a single, unified system -- a gap that \textit{ChemReporter} directly addresses.

% \subsection{Data Anomaly Mitigation and Active Curation}\todo{is this enough ? i describe more later  }

% Each datasets has come out with specialised methods to curate their  data especially OC20. To address dataset redundancy and domain representative bias, recent techniques have advanced automated data sub-selection using probabilistic frameworks like Determinantal Point Processes (DPPs) to enforce diversity across configuration sets \cite{dpp_mlip_2026}. While DPPs effectively sample structural space, they scale poorly and lack the capacity to explicitly detect physical corruption. 

\section{Design Principles and Workflow}

\begin{figure}[htbp]
\centering
\includegraphics[width=\textwidth]{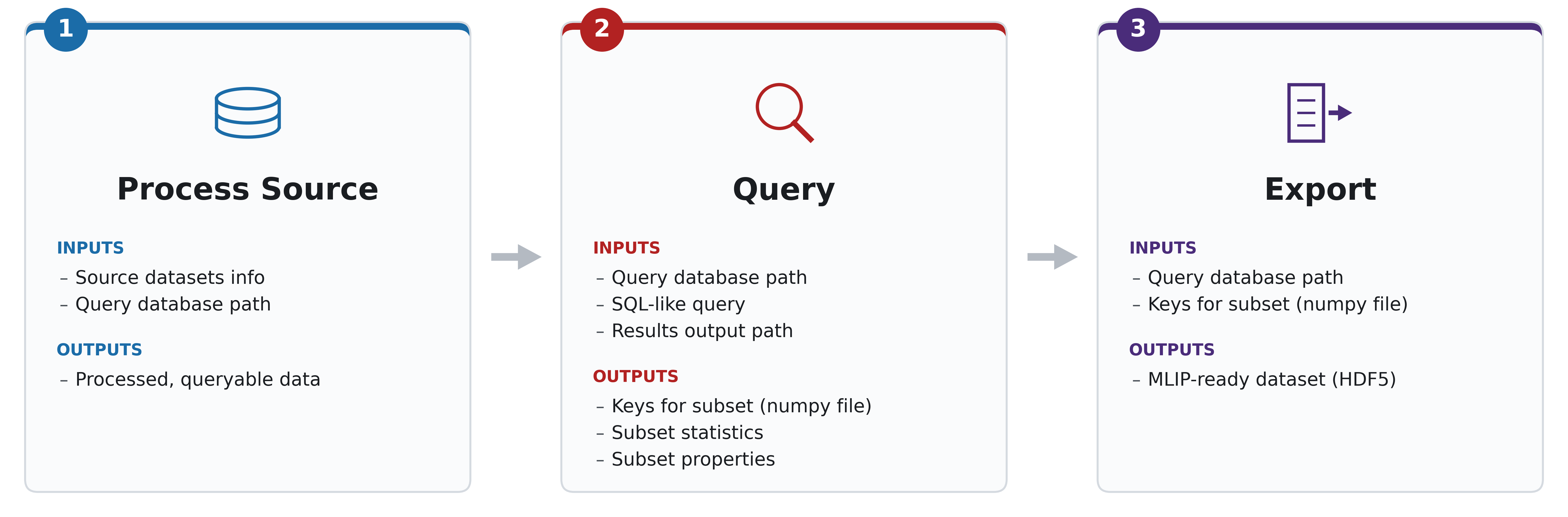}
\caption{Overview of the \textit{ChemReporter} pipeline, consisting of three decoupled stages: (1) \textit{processing}, which converts a raw source database into a structured, queryable Parquet repository; (2) \textit{querying}, which filters and samples this repository using SQL-like criteria to produce a subset of entry keys along with associated statistics; and (3) \textit{exporting}, which uses these keys to generate an HDF5 dataset ready for training of MLIPs. Each stage can be run independently, using the outputs of the previous stage as its inputs.}
\label{fig:workflow}
\end{figure}

The architecture of \textit{ChemReporter} is built upon three pillars: modularity, scalability, and reproducibility. Traditional workflows in computational chemistry are often slowed down by fragmented, one-off scripts. To resolve these bottlenecks, \textit{ChemReporter} separates data preparation into clear, independent steps for (pre-)processing, querying, and exporting, as illustrated in the overall workflow (Figure~\ref{fig:workflow}). This design allows users to run each part of the pipeline independently, for example, preprocessing a new dataset once, then querying it repeatedly to refine a subset, before exporting only the final result. During preprocessing, \textit{ChemReporter} ingests large raw chemical datasets, extracts core structural and quantum chemical properties, and stores the result in a database of partitioned Apache Parquet files. For querying, the framework leverages Polars\footnote{See \url{https://pola.rs/}.} to perform out-of-core, 
streaming DataFrame operations across these partitioned files, lazily evaluating user queries and filtering millions of structures chunk-by-chunk with a minimal, constant memory footprint. During export, \textit{ChemReporter} streams the selected configurations and writes them in parallel across multiple worker threads into clean HDF5 files, fully prepared for downstream machine learning training.
We describe each of these three stages (processing, querying, and exporting) in more detail in the subsections below.

\subsection{Processing}

During processing, \textit{ChemReporter} parses raw atomic configurations, i.e., atomic coordinates, elements, and associated level-of-theory information, into structured Polars DataFrames. From this data, it computes essential physical properties (e.g., net forces, dipole moments, molecular weights) and applies a set of sanity-based filtering criteria, such as bond-length and connectivity checks, or maximum force thresholds, to flag or exclude invalid molecular configurations. For broader chemical analysis, the pipeline can optionally compute topological graph properties and molecular fingerprints using RDKit, allowing structures to be subsequently filtered or selected based on specific fingerprint types.

Processing is run via the \textit{ChemReporter} CLI, for example using the command \texttt{chemreporter process -{}-config process.yaml}. The configuration file (\texttt{process.yaml}) allows users to specify processing chunk sizes, toggle graph-based feature extraction such as fingerprint computation, and adjust general processing settings, for instance, imposing atom-count limits or excluding specific data subsets -- to manage memory and compute usage.

In addition to per-structure properties, \textit{ChemReporter} also records reaction-level metadata, such as reactant/product/transition-state labels and reaction energy barriers, for datasets that are organized around chemical reactions rather than isolated structures. Table~\ref{tab:chemreporter_fields} summarizes the full set of fields extracted and computed during processing, spanning metadata, physical properties, structural descriptors, and reaction-level information.

\begin{table}[h]
\centering
\caption{Summary of key fields extracted and computed by the \textit{ChemReporter} processing pipeline.}
\label{tab:chemreporter_fields}
\vspace{0.3cm}
\renewcommand{\arraystretch}{1.3}
\setlength{\tabcolsep}{6pt}
\begin{tabular}{l>{\raggedright\arraybackslash}p{9cm}}
\toprule
\textbf{Category} & \textbf{Fields} \\
\midrule
Basic Metadata & \texttt{entry\_key}, \texttt{database\_name}, \texttt{split\_name}, \texttt{subset}, \texttt{source} \\
DFT Metadata & \texttt{basis\_set}, \texttt{functional}, \texttt{correction\_term} \\
Physical Properties & \texttt{energy}, \texttt{net\_charge}, \texttt{spin\_multiplicity}, \texttt{dipole\_moment\_magnitude} \\
Atomic Information & \texttt{composition}, \texttt{num\_atoms}, \texttt{atomic\_numbers}, \texttt{atomic\_symbols}, \texttt{molecular\_weight} \\
Force Metrics & \texttt{net\_force\_norm}, \texttt{max\_force\_norm} \\
Molecular Descriptors & \texttt{logp}, \texttt{tpsa}, \texttt{smiles} \\
Fingerprints & \texttt{fingerprint\_0}, \ldots, \texttt{fingerprint\_1023} \\
Reaction Data & \texttt{reaction\_id}, \texttt{is\_reactant}, \texttt{is\_product}, \texttt{is\_transition\_state}, \texttt{energy\_barrier} \\
Structural Checks & \texttt{is\_molecular\_structure\_valid}, \texttt{num\_water\_molecules}, \texttt{is\_protein}, \texttt{is\_nucleobase} \\
\bottomrule
\end{tabular}
\end{table}

% --- OLD TABLE: ---
% \begin{table}[h]
% \centering
% \begin{tabular}{|l|l|}
% \hline
% \textbf{Category} & \textbf{Fields} \\ \hline
% Base Metadata & \texttt{entry\_key}, \texttt{database\_name}, \texttt{split\_name}, \texttt{subset}, \texttt{source} \\ \hline
% DFT Metadata & \texttt{basis\_set}, \texttt{functional}, \texttt{correction\_term} \\ \hline
% Physical Properties & \texttt{energy}, \texttt{net\_charge}, \texttt{spin\_multiplicity}, \texttt{dipole\_moment\_magnitude} \\ \hline
% Atomic Information & \texttt{composition}, \texttt{num\_atoms}, \texttt{atomic\_numbers}, \texttt{atomic\_symbols}, \texttt{molecular\_weight} \\ \hline
% Force Metrics & \texttt{net\_force\_norm}, \texttt{max\_force\_norm} \\ \hline
% Graph Properties & \texttt{logp}, \texttt{tpsa}, \texttt{smiles}, \texttt{fingerprint\_0} to \texttt{fingerprint\_1023} \\ \hline
% Reaction Data & \texttt{reaction\_id}, \texttt{is\_reactant}, \texttt{is\_product}, \texttt{is\_transition\_state}, \texttt{energy\_barrier} \\ \hline
% Structural Checks & \texttt{is\_molecular\_structure\_valid}, \texttt{num\_water\_molecules}, \texttt{is\_protein}, \texttt{is\_nucleobase} \\ \hline
% \end{tabular}
% \caption{Summary of key fields extracted and computed by the \textit{chemreporter} processing pipeline.}
% \label{tab:chemreporter_fields}
% \end{table}

\subsection{Querying}

\subsubsection{Command-Line Interface}

The CLI allows users to create a filtered subset of a preprocessed chemical dataset by specifying selection criteria in a configuration file. Users can initiate the querying and filtering pipeline directly from the terminal via a single command, \texttt{chemreporter query -{}-config query.yaml}, which points to a configuration file specifying the input and output data paths, the query itself, and any sampling parameters.

% \begin{verbatim}
% # Example: Processing a raw database via the CLI
% chemreporter process --config process_db.yaml
% \end{verbatim}

% \begin{verbatim}
% # process_db.yaml
% query_database_path: "s3://path-to-the-query-db"
% results_path: "s3://result_path"

% query: "database_name = 'oc20' AND net_force_norm < 1.0e-3 AND max_force_norm < 1.5"

% sample_actions:
%   n_samples: 100000
%   sampling_method: random
%   seed: 42
% \end{verbatim}

The query is expressed as a logical filter expression over the fields described in Table~\ref{tab:chemreporter_fields}, allowing users to combine dataset-level filters (e.g., selecting a specific source dataset) with physical or structural constraints. For example, a user might filter for stable configurations by combining a near-zero net-force condition with an upper bound on the maximum atomic force, isolating structures that are both close to equilibrium and free of unphysically large force outliers. In addition to filtering, the configuration file allows users to specify a sampling strategy, for instance, drawing a fixed-size random subsample with a defined random seed to ensure reproducibility, to further reduce the filtered result to a manageable, representative subset. Moreover, custom sampling functions can be integrated easily by the user.

By default, \textit{ChemReporter} reads from and writes to local file paths, but users can readily supply their own custom read and upload functions to interface with remote or cloud storage on either end, retrieving the input dataset to be queried, and transferring the resulting subset elsewhere, allowing the tool to be integrated into existing cloud-based data pipelines without native cloud support being built into the core querying step.

% As demonstrated in the configuration example, \textit{ChemReporter} natively supports cloud storage backends like AWS S3, allowing users to point directly to remote data repositories (\texttt{query\_database\_path}) without manual local downloads. This configuration filters the \texttt{oc20} dataset for stable configurations based on strict physical constraints, specifically isolating structures with near-zero net forces (\texttt{net\_force\_norm < 1.0e-3}) and uniform individual atomic forces (\texttt{max\_force\_norm < 1.5}). Under the \texttt{sample\_actions} block, the tool applies a reproducible, random sampling algorithm to extract exactly 100,000 target structures, exporting the final database entry keys directly back to cloud storage as a NumPy array (\texttt{results\_path}).

\subsubsection{Python API}

For more flexibility, users can interactively explore, manipulate, and filter the data using the Python API. At its core is a class, the \texttt{QueryDatabaseHandler}, which serves as the programmatic entry point to a processed dataset, exposing its contents as a Polars DataFrame. This gives users direct access to the full expressiveness of Polars for custom analyses, such as computing reaction energy barriers, isolating specific transition states, or applying arbitrary custom filters, without needing to load the entire dataset into memory.

% \begin{verbatim}
% import polars as pl
% from chemreporter.query_database_tools.query_database import QueryDatabaseHandler

% # Initialize the handler with the processed database repository
% handler = QueryDatabaseHandler(db_path="path/to/processed_db")

% # Filter for accessible reactions and locate transition states 
% # based on the highest energy point along the trajectory
% results = handler.query(
%    "energy_barrier" < 2.0],
% )

% # Generate stratified train/val/test splits balanced across 
% # distinct chemical reaction types
% splits = handler.generate_splits(
%     ratios={"train": 0.8, "val": 0.1, "test": 0.1}, by="smiles"
% )  
% \end{verbatim}
% \todo{Actually implement the splits or remove the last example  OR use the correct syntax from the tutorial notbook ;-) }

\subsection{Exporting}

Once the desired entry keys have been generated (either via the CLI or the Python API), \textit{ChemReporter} uses them to extract and structure the corresponding configurations into a machine-learning-ready HDF5 file. During this export phase, \textit{ChemReporter} maps the keys back to the underlying partitioned database and streams only the requested configurations, chunk-by-chunk. To bypass traditional single-file serialization bottlenecks, the framework processes these chunks in parallel across multiple worker threads. The resulting HDF5 file stores atomic numbers, coordinates, forces, energies, periodic boundary conditions, charge and spin multiplicity, along with any other optionally requested properties relevant to the downstream application. This design allows the output file to be streamed directly into modern MLIP training frameworks, such as the \textit{mlip} library~\cite{brunken2026machine}, with no additional data conversion required.

\section{Computational Efficiency and Scalability}
\label{sec:benchmarks}

Dataset curation for MLIP training rarely happens in a single pass: constructing a well-curated training set typically involves processing millions to billions of raw structures, often across multiple rounds of filtering. As dataset sizes grow, naively loading such data into memory becomes infeasible on standard hardware, and inefficient processing can turn a routine curation step into a multi-hour bottleneck. Computational efficiency is therefore a practical requirement for making large-scale, iterative dataset curation feasible at all.

\begin{figure}[htbp]
\centering
\includegraphics[width=0.8\textwidth]{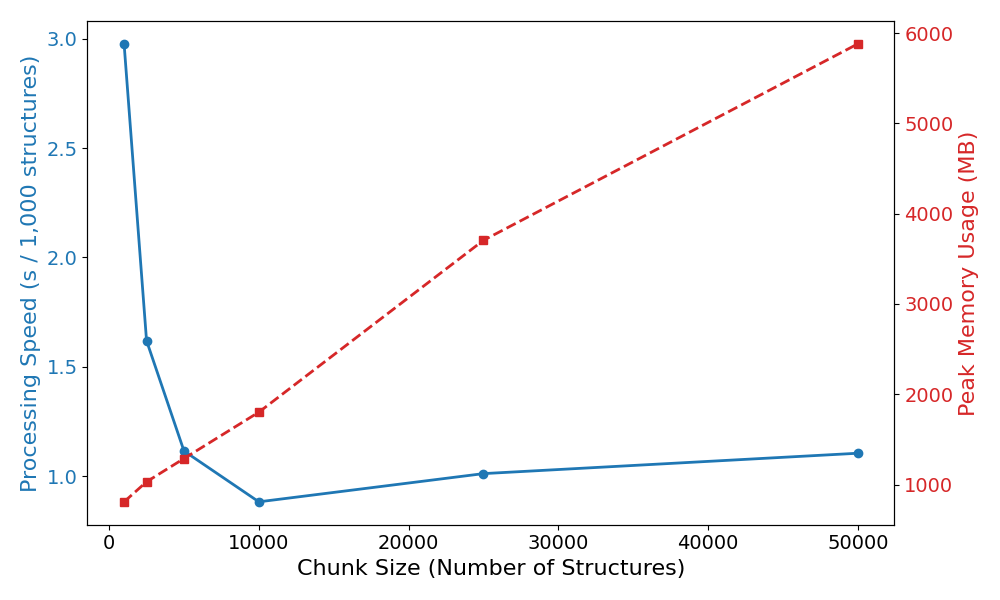}
\caption{Processing speed and peak memory usage as a function of chunk size with graph-based properties disabled. Larger chunk sizes incrementally increase peak memory utilization while stabilizing processing speed.}
\label{fig:chunk_graph_off}
\end{figure}

To evaluate the computational efficiency and scalability of \textit{ChemReporter}, we benchmarked processing speed and memory utilization across a range of conditions. The most computationally intensive phase of the pipeline involves parsing raw atomic structures (via ASE) into structured DataFrames, computing physical properties, and optionally generating graph-derived features via RDKit. Processing massive chemical datasets often presents significant memory bottlenecks; to mitigate this, \textit{ChemReporter} employs a chunked processing architecture backed by high-performance \texttt{Polars} DataFrames.
Pipeline performance depends on the number of structures processed per chunk. We measured processing speed (seconds per 1,000 structures) and peak memory usage across chunk sizes ranging from 1,000 to 50,000 structures, both with and without extended graph-based feature extraction enabled. As shown in Figure~\ref{fig:chunk_graph_off}, when graph-based processing is disabled, the pipeline achieves high throughput; increasing the chunk size predictably increases peak memory footprint but yields diminishing speed gains once I/O overhead is amortized, allowing users to tune chunk size to their available hardware (e.g., laptops versus high-memory compute nodes) without meaningfully sacrificing processing speed.

\begin{figure}[htbp]
\centering
\includegraphics[width=0.8\textwidth]{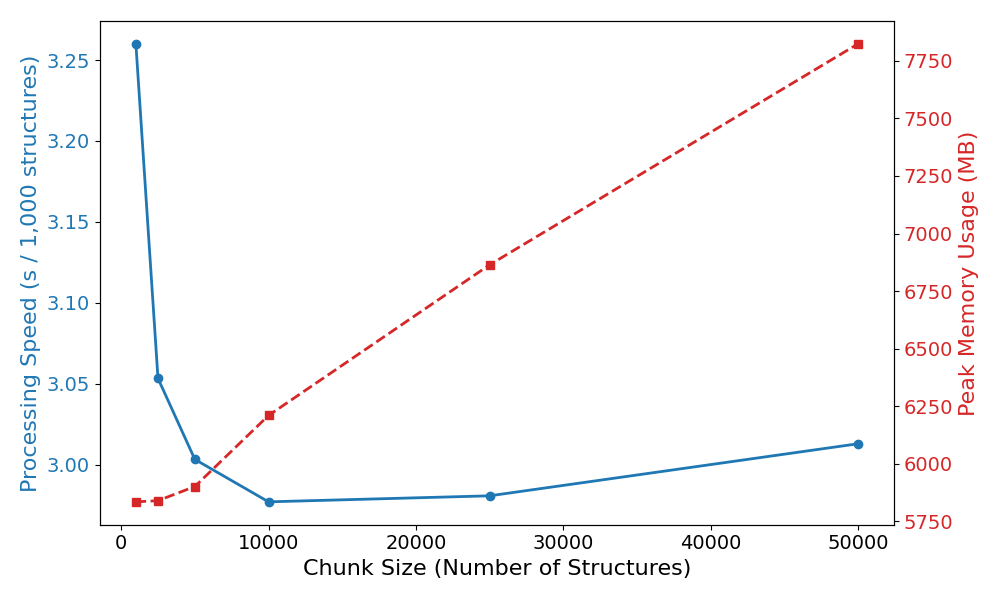}
\caption{Processing speed and peak memory usage with graph-based processing enabled. Despite the increased computational complexity of RDKit property generation, memory usage remains well-bounded through efficient chunking.}
\label{fig:chunk_graph_on}
\end{figure}

When advanced graph-based topological properties (via RDKit) are enabled (Figure~\ref{fig:chunk_graph_on}), peak memory and computational requirements increase accordingly; however, the chunking mechanism keeps memory growth linear and bounded, preventing out-of-memory failures even when computing complex fingerprints for dense datasets.

\begin{figure}[htbp]
\centering
\includegraphics[width=.8\textwidth]{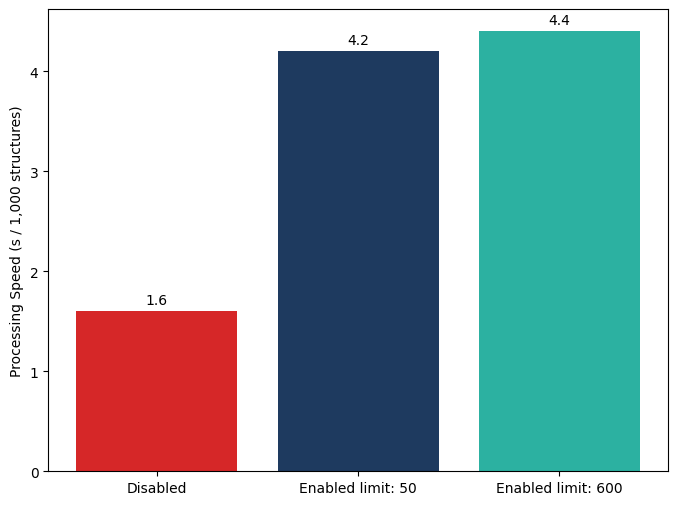}
\caption{Computational overhead comparison between baseline processing and extended graph property extraction. Enabling graph-based feature extraction introduces most of the added cost; the atom-count limit provides a secondary lever for capping overhead on occasional very large structures without materially affecting throughput otherwise.}
\label{fig:graph_overhead}
\end{figure}

\textit{ChemReporter} also allows users to toggle graph-based feature extraction entirely or restrict it by molecular size, bounding the overhead introduced by RDKit property generation relative to baseline structural processing (see Figure~\ref{fig:graph_overhead}). This allows occasional very large structures (e.g., bulk solvent boxes or periodic slabs) to be excluded from graph extraction so they do not stall the pipeline.

Overall, these results demonstrate that \textit{ChemReporter} achieves efficient, tunable dataset processing suitable for both rapid prototyping and large-scale dataset curation.

\section{Queryable Chemical and Physical Properties}
\label{sec:properties}

During processing, \textit{ChemReporter} computes a broad set of chemical and physical properties for each structure, which are stored alongside the raw data and can subsequently be used as query criteria (see Table~\ref{tab:chemreporter_fields}). These properties fall into four broad categories, described below.

\begin{figure}[t]
    \centering
    
        \centering
        \includegraphics[width=.8\textwidth]{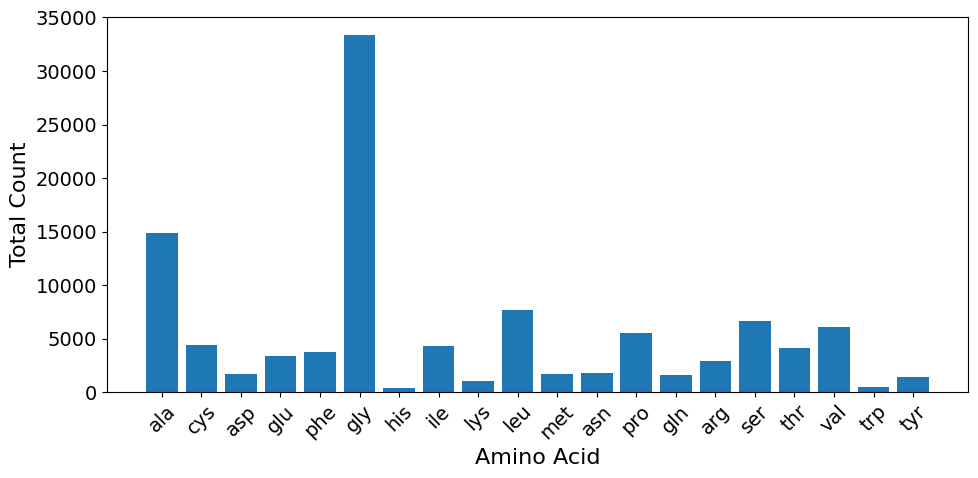}
        \caption{Distribution of standard amino acid residues identified by \textit{ChemReporter}'s biological motif detection in the biomolecular subset of OMol25. Each structure is scanned for standard amino acid substructures, and the resulting per-residue counts are stored as queryable metadata, illustrating how \textit{ChemReporter} enables biochemistry-specific dataset characterization beyond generic structural properties.}
        \label{fig:aminoacids}

\end{figure}

\begin{figure}[t]
    \centering

    \begin{subfigure}{0.48\textwidth}
        \centering
        \includegraphics[width=\linewidth]{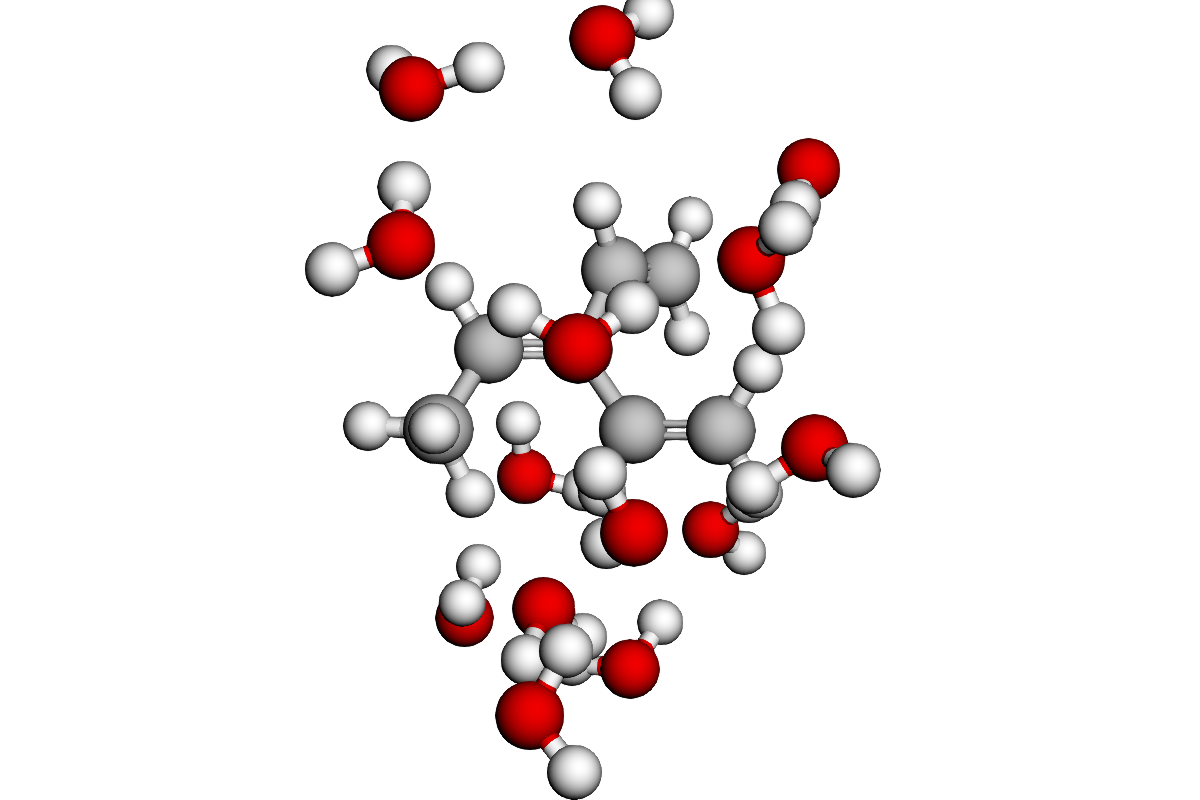}
        \caption{}
        \label{fig:biofirst}
    \end{subfigure}
    \hfill
    \begin{subfigure}{0.48\textwidth}
        \centering
        \includegraphics[width=\linewidth]{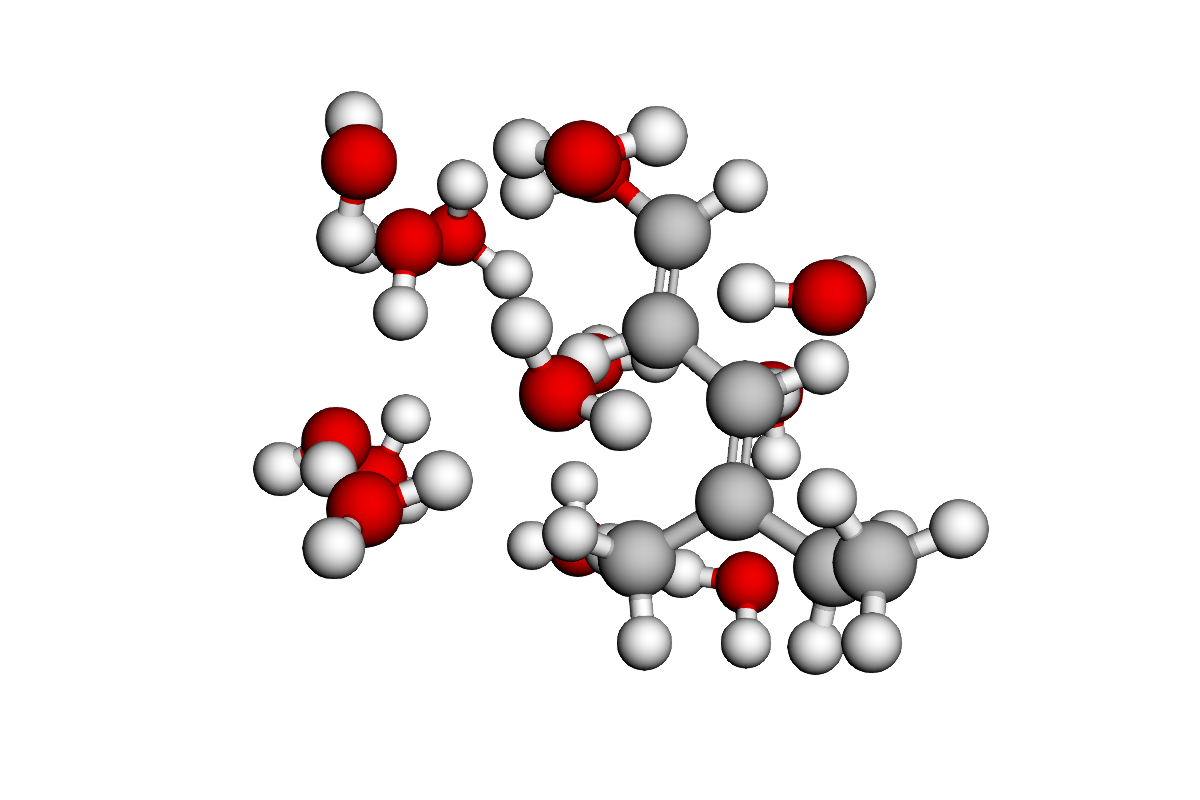}
        \caption{}
        \label{fig:biosecond}
    \end{subfigure}

    \caption{Examples of structures flagged and excluded during biomolecular dataset curation with \textit{ChemReporter}. (a) A branched carbohydrate chain and (b) a conjugated alkene system, both present in the OMol25 biomolecular subset, are correctly identified as not corresponding to standard biological motifs (e.g., amino acid residues or nucleobases) and are tagged accordingly.}
    \label{fig:biomotifs}
\end{figure}

\begin{enumerate}
    \item \textbf{Topological Representations:}
    \begin{itemize}
        \item \textit{Canonical SMILES:} Simplified Molecular Input Line Entry System (SMILES) strings are generated from 3D atomic coordinates and atomic numbers using open-source cheminformatics toolkits such as OpenBabel~\cite{oboyle2011open}. These pipelines include fallback sanitization routines to handle complex geometries, such as highly strained macrocycles or organometallic complexes.
        \item \textit{Topological Fingerprints:} Standard 1024-bit Morgan fingerprints are generated for each valid molecular configuration using the extended-connectivity circular fingerprint (ECFP) formalism~\cite{rogers2010extended}, encoding the local chemical environment of each atom up to a specified radius. These fingerprints enable ultra-fast substructure filtering and Tanimoto similarity queries directly within the data backend, without needing to parse raw 3D structural files.
    \end{itemize}

    \item \textbf{Physicochemical Descriptors:}
    The partition coefficient (LogP) and Topological Polar Surface Area (TPSA) are computed via RDKit~\cite{landrum2016rdkit}, providing physicochemical properties relevant for filtering structures in targeted molecular design and virtual screening applications.

    \item \textbf{Biological Motif Counts:}
    For biomolecular structures, \textit{ChemReporter} applies substructure matching to identify and count standard biological building blocks, such as individual amino acid residues and solvent (e.g., water) molecules. This semantic metadata makes the resulting dataset directly queryable for biochemistry-focused applications, without requiring users to re-derive this information themselves. This process yields the amino acid composition shown in Figure~\ref{fig:aminoacids} and successfully identifies biologically implausible motifs (see Figure~\ref{fig:biomotifs}).

    \item \textbf{Geometric and Physical Properties:}
    For all structures, including extended solid-state systems where molecular graph definitions are not well defined, \textit{ChemReporter} computes global physical invariants such as the net force norm, maximum atomic force norm, dipole moment magnitude, and total atom count.
\end{enumerate}

\section{Dataset Curation with ChemReporter} \label{sec:curation}

To train an accurate MLIP model, dataset sanitation plays a central role in model reliability~\cite{creed2026openquestionsmachinelearnedinteratomic}, and curation must address both dataset-agnostic structural pathologies and anomalies specific to individual source datasets~\cite{Kulichenko2024}. Training on uncurated datasets forces the neural network to approximate non-physical discontinuities, fundamentally warping the learned Potential Energy Surface (PES) and amplifying downstream distribution shifts during molecular dynamics simulations~\cite{KreimanKrishnapriyan2025}.

It is important to state a strict caveat on the intended use of dataset curation: it must not be used to artificially inflate apparent model performance by removing structures a model simply finds difficult. Legitimate curation serves only two purposes: (1) identifying and removing genuinely poor-quality or unphysical data, as described above, and (2) deliberately selecting and documenting, with full traceability, a constrained training distribution for a model with intentionally limited expressivity and a correspondingly narrow application scope. Any use of curation beyond these two purposes risks producing misleadingly optimistic performance metrics that do not reflect a model's true reliability.

\subsection{General Structural Validation} \label{sec:struct_validation}

Structural anomalies can arise even when the underlying reference \textit{ab initio} method (typically Density Functional Theory) completes without a system crash, yet yields a mathematically corrupt description of the PES~\cite{Kulichenko2024}. For instance, the Self-Consistent Field solver can settle into a metastable local electronic minimum rather than converging to the true electronic ground state, creating localized energy spikes ($\Delta E > 2\text{ eV}$) for geometries that are otherwise structurally sound. Separately, automated high-throughput generation pipelines (e.g., high-temperature \textit{ab initio} molecular dynamics or active learning sampling) can introduce geometric inconsistencies such as steric clashes and core overlap~\cite{Kulichenko2024}: when interatomic distances fall far below the sum of their covalent radii ($d_{ij} \ll r_i + r_j$), the resulting nuclear repulsion can produce forces exceeding $100\text{ eV/\AA}$, a well-documented cause of gradient explosions that can permanently disrupt a neural network's weights during training.

To catch these and related failure modes, \textit{ChemReporter} applies three general, dataset-agnostic structural checks. \textbf{Topological and valency validation} parses atomic connectivity to ensure it respects standard chemical valency constraints, flagging configurations with chemically implausible bonding patterns (e.g., a carbon atom bonded to five or more distinct atoms). \textbf{Graph connectivity scans} analyze molecular graphs to detect fragmented or discontinuous structures, identifying conformations containing isolated or floating atoms not properly integrated into the molecular framework. \textbf{Spatial and distance thresholding} evaluates atomic coordinate matrices to flag atom pairs separated by distances below a defined fraction of the sum of their covalent radii ($r_{ij} < \text{threshold}$), directly targeting the steric-clash failure mode described above. Together, these checks provide a general-purpose structural screening layer, removing configurations that could otherwise introduce large, destabilizing force vectors into MLIP training pipelines.

\subsection{Dataset-Specific Curation}

Because different datasets carry unique structural properties and generation errors, \textit{ChemReporter}'s filtering pipeline also supports criteria tailored to specific source datasets, applied in addition to the general checks described above.

\subsubsection{Curation of Biomolecular Systems}

Automated preparation pipelines introduce distinctive structural defects during biomolecular dataset generation. Three main factors drive these artifacts: the high flexibility of macromolecular degrees of freedom, intricate non-covalent interactions, and the constrained system sizes imposed to maintain efficient processing during MLIP training. As a result, these datasets frequently contain corrupted structural entries or unphysical geometries. Within macromolecular subsets such as those found in OMol25, \textit{ChemReporter} applies the same general valency, connectivity, and clash-detection checks described above, which are particularly effective at catching the flexible-chain and non-covalent-interaction artifacts characteristic of this domain.

\subsubsection{Curation of Catalyst Datasets}  

\begin{figure}[htbp]
    \centering

    \begin{subfigure}[b]{0.30\textwidth}
        \centering
        \includegraphics[width=\textwidth]{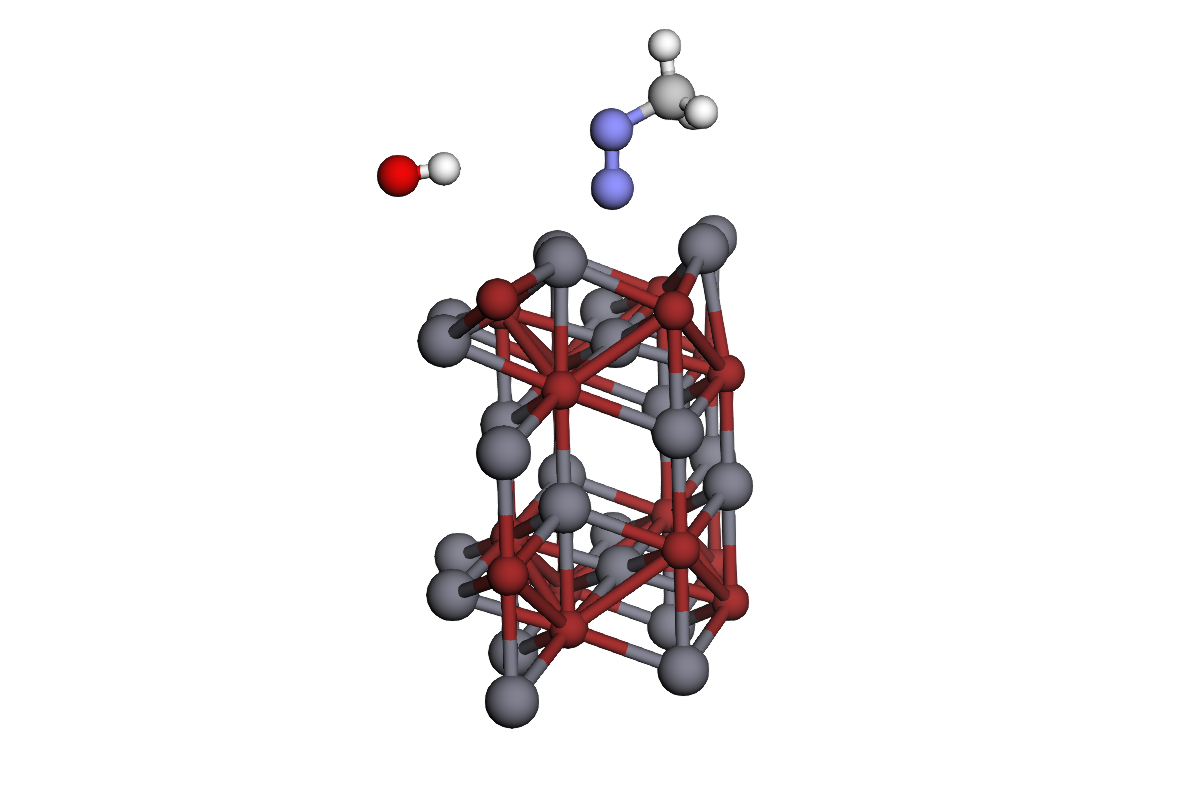}
        \caption{}
        \label{fig:img1}
    \end{subfigure}
    \hfill
    \begin{subfigure}[b]{0.30\textwidth}
        \centering
        \includegraphics[width=\textwidth]{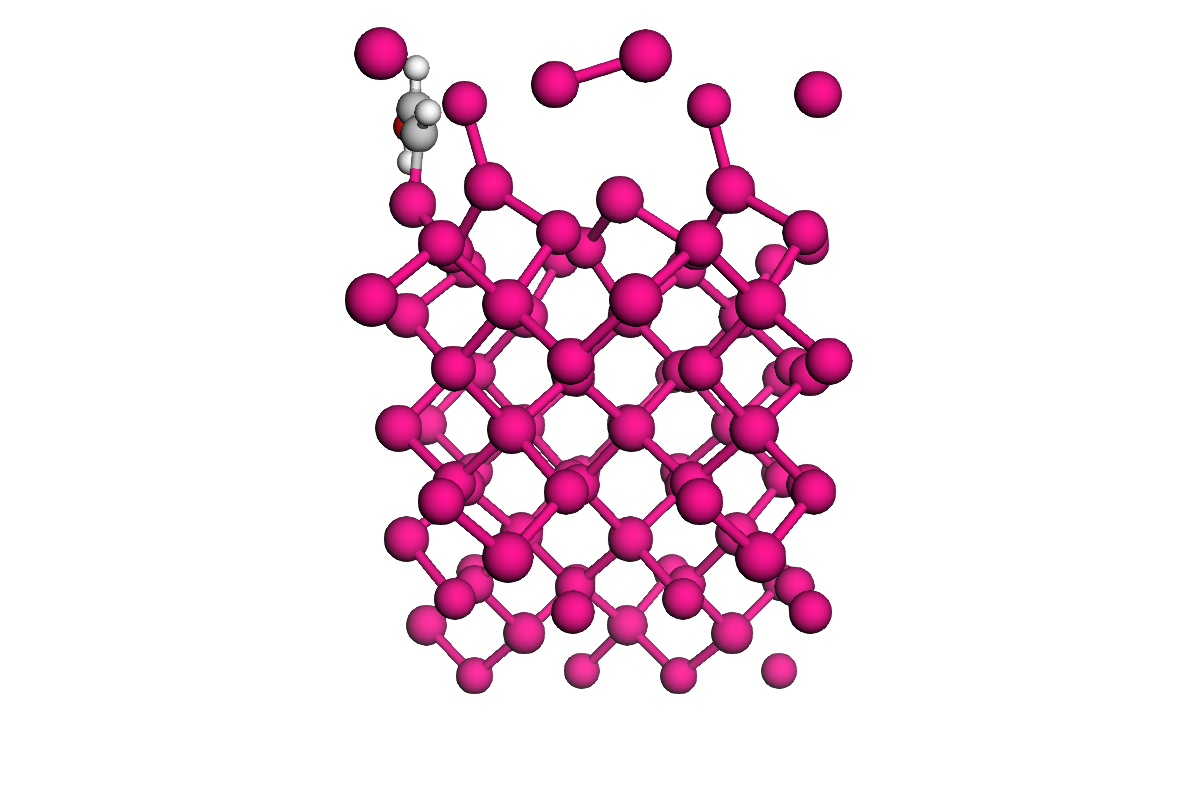}
        \caption{}
        \label{fig:img2}
    \end{subfigure}
    \hfill
    \begin{subfigure}[b]{0.30\textwidth}
        \centering
        \includegraphics[width=\textwidth]{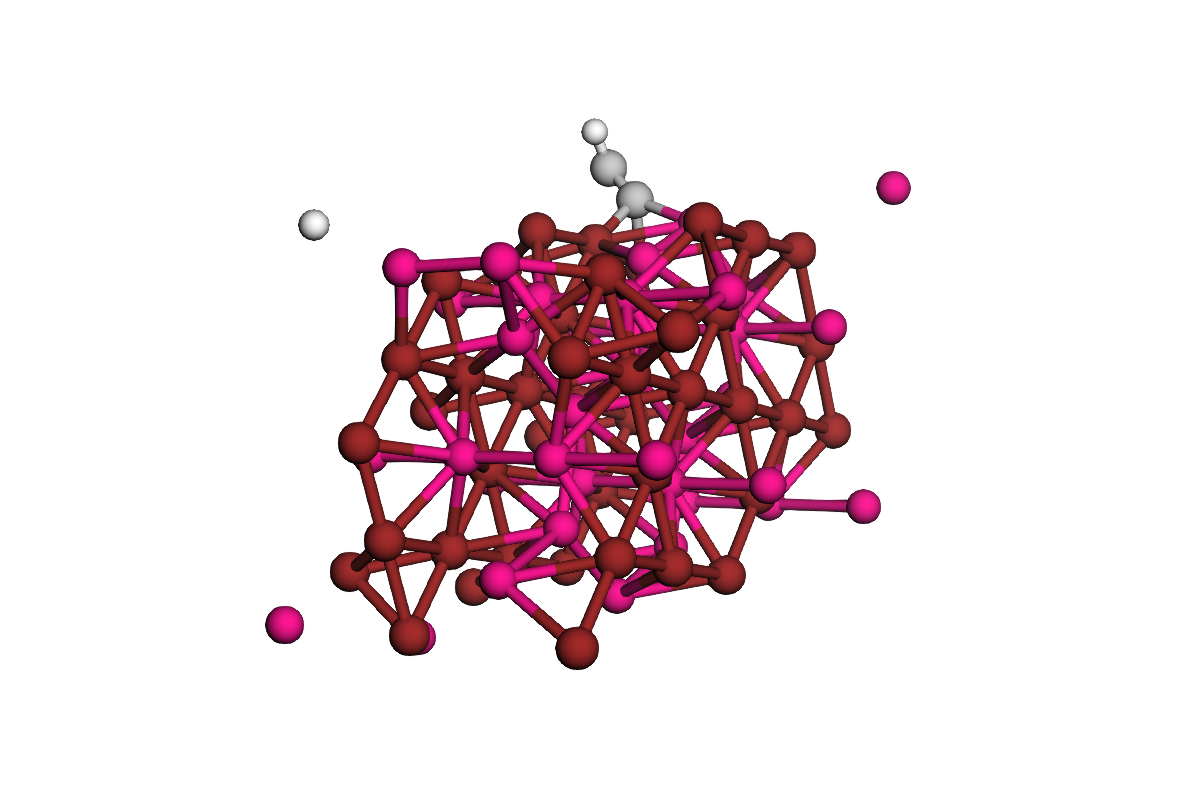}
        \caption{}
        \label{fig:img3}
    \end{subfigure}

    \vspace{0.5cm} % Space between rows

    \caption{Representative examples of unphysical structural anomalies flagged and filtered by the catalyst curation pipeline: (a) the substrate is split, with part of it floating in vacuum; (b) the slab lattice is structurally compromised; (c) the structure contains floating atoms.}
    \label{fig:oc20_grid}
\end{figure}

Catalyst-adsorbate systems are particularly prone to unphysical or non-equilibrium geometries arising during high-throughput structural relaxations. To prevent such anomalies from introducing noise or training instabilities, \textit{ChemReporter}'s catalyst-specific pipeline screens for three primary types of structural artifacts. First, adsorbates placed too close to multiple surface atoms simultaneously can create hypervalent coordination states that do not exist in nature. Second, adsorbates may desorb from the catalyst surface during relaxation, crossing the vacuum layer boundary; such configurations are flagged when adsorbate atoms sink below the first catalyst layer or cross the simulation cell boundaries. Third, a spatial filter detects fractured slab surfaces, where a slab atom dissociates completely from the lattice, leaving an unphysical lone atom suspended in the vacuum. Figure~\ref{fig:oc20_grid} shows representative examples of each of these failure modes.

\subsection{Curation Impact on Training Performance}

Having described \textit{ChemReporter}'s curation capabilities in the previous section, we now present a controlled experiment demonstrating their practical impact on MLIP training. Our goal is to show, using a simple and reproducible setup, that even light-touch curation (e.g., removing a small fraction of structurally or physically anomalous configurations) can measurably improve training outcomes, rather than merely serving as a theoretical safeguard.

\subsubsection{Experimental Setup} \label{sec:exp_setup}

We use the neutral subset of SPICE2 structures within OMol25, which after splitting yields approximately 1.6 million training structures. From this pool, we construct two versions of the training set: an \emph{uncurated} version containing all available structures, and a \emph{curated} version obtained by applying three filtering criteria via \textit{ChemReporter}: (i) a near-zero net force condition (the norm of the summed force vector below $10^{-3}$~eV/\AA), (ii) an upper bound on local force outliers (maximum atomic force below 15~eV/\AA), and (iii) chemical validity according to our connectivity-based molecular structure checks (see section~\ref{sec:struct_validation}). Together, these criteria remove approximately 6,000 structures, or under 0.4\% of the original training pool, a deliberately small fraction, chosen to test whether even minimal, targeted curation yields a measurable training benefit, rather than relying on aggressive filtering to make the effect obvious, which would risk discarding high-energy structures that are physically accessible during application and simply hard to predict, rather than genuinely invalid.

For validation, we use 100,000 structures drawn exclusively from the curated pool, in both settings. This choice reflects the intended use case for a trained MLIP: at deployment or in downstream simulation, a model is expected to encounter physically reasonable configurations, not the unphysical or numerically pathological structures that curation is designed to catch. Such structures would typically cause a simulation to fail outright, independent of model quality, rather than reflecting a meaningful test of predictive accuracy. Evaluating exclusively on curated validation data therefore lets us isolate a specific question: does including the additional, anomalous training structures actually help the model, or does it degrade performance on the realistic structures it is actually meant to handle? Using an uncurated validation set would conflate these two effects, since any apparent benefit or harm from the anomalous training examples could be masked or exaggerated by the model's behavior on anomalous validation examples as well. 

We note that our filtering thresholds target only structures with severe, well-characterized physical pathologies (e.g., force values far outside stable, near-equilibrium ranges, or chemically inadmissible connectivity), which are not representative of configurations encountered during ordinary application. We therefore do not claim curation improves performance on atypical-but-valid, high-energy structures; rather, we show that removing genuinely unphysical entries carries no cost on realistic validation data, and in fact yields a measurable benefit.

We train and evaluate two model architectures, MACE and ViSNet, on both the curated and uncurated training sets, with two independent initialization seeds per model/dataset combination to account for variance due to random weight initialization. The complete training and architecture details for each model are provided in Appendix~\ref{app:mace_training} and Appendix~\ref{app:visnet_training}, respectively.

\subsubsection{Results}

% On the validation set, training on the filtered data produced a small but consistent improvement. Averaged across architectures, the energy MAE decreased from $1.02 \pm 0.08$ to $0.98 \pm 0.16$~meV/atom, while the force MAE decreased from $17.64 \pm 0.29$ to $16.87 \pm 0.39$~meV/\AA. Training dynamics were essentially unchanged, and the training converged over the same number of epoch, with a marginal gain in wall-clock time. 
% Overall, the effect of filtering is small but consistently positive. The reduction in both energy and force errors indicates that improving training-data quality is more beneficial than retaining a small number of potentially noisy structures, while having negligible impact on training efficiency.

\begin{figure}[h!]
  \centering
  \includegraphics[width=0.85\linewidth]{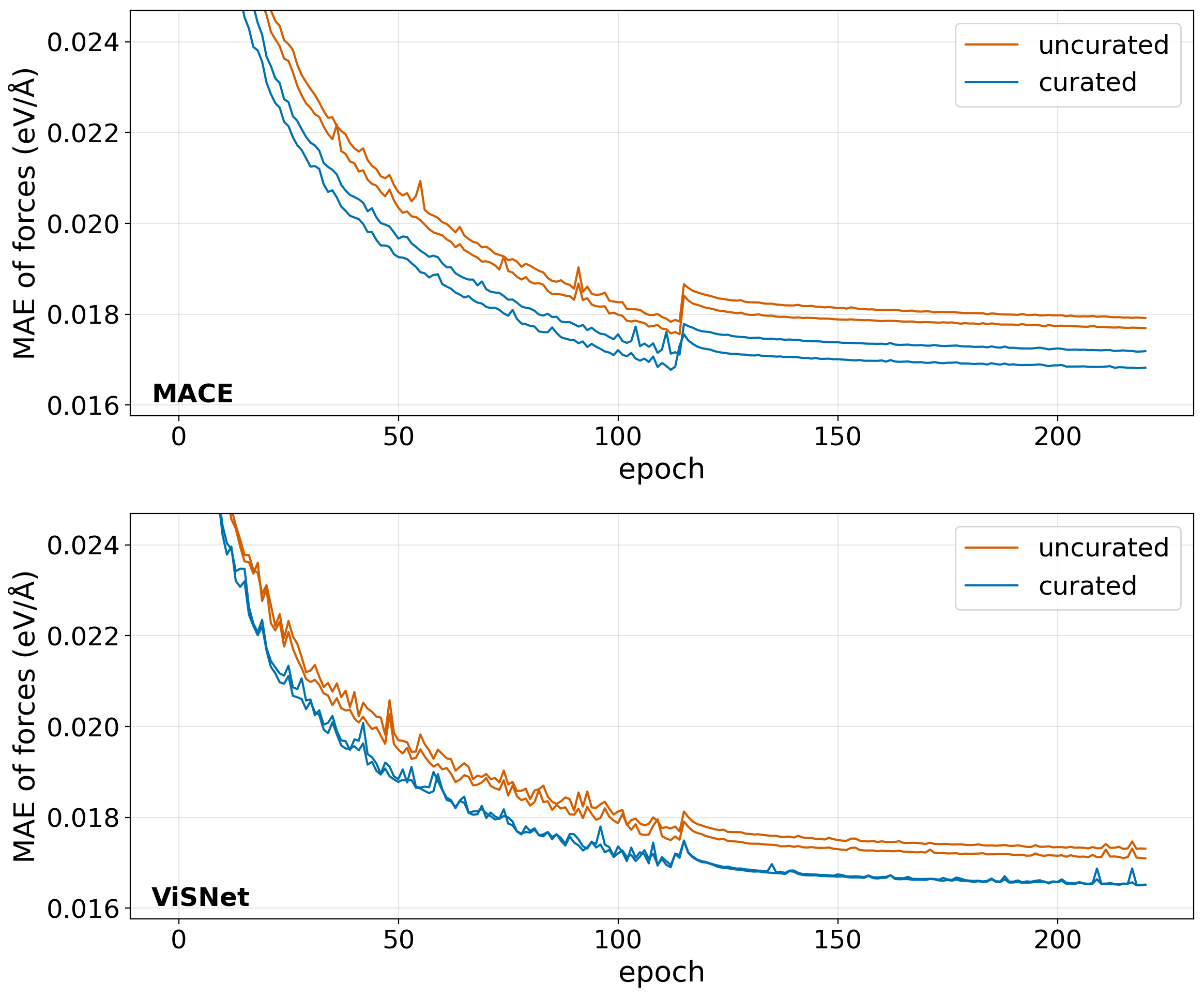}
\caption{Validation force MAE as a function of training epoch for MACE (top) and ViSNet (bottom), comparing models trained on curated vs.\ uncurated SPICE2. Across both architectures, models trained on the curated dataset achieve consistently lower validation force error throughout training. The visible discontinuity around epoch 115 corresponds to the loss weight-flipping schedule. Two lines per dataset correspond to results from two independent initialization seeds.}
  \label{fig:curated-vs-uncurated-curve-force-mae}
\end{figure}

On the validation set, training on the curated data produced a small but consistent improvement
in force accuracy for both architectures: force MAE decreased from 17.9 to 17.2\,meV/Å for MACE
and from 17.3 to 16.5\,meV/Å for ViSNet, with force RMSE improving similarly (see Table~\ref{tab:force-curated-vs-uncurated-mlflow}). Figure~\ref{fig:curated-vs-uncurated-curve-force-mae}
shows this improvement is consistent throughout training, with the curated models tracking below the
uncurated ones from early epochs onward for both architectures.

For energy errors, the effect of curation was more mixed. Energy MAE improved slightly for MACE
(33.1 to 33.0\,meV) and more substantially for ViSNet (26.9 to 24.4\,meV) (see Table~\ref{tab:energy-curated-vs-uncurated-mlflow}). Energy RMSE
followed the same trend for ViSNet (51.0 to 46.2\,meV), but for MACE, curation slightly increased
energy RMSE (59.7 to 62.0\,meV), the one deviation from an otherwise consistent trend across both
metrics and architectures. Corresponding training curves for energy loss and MAE are provided in
Appendix~\ref{app:training_curves}.

Overall, curation improves force prediction accuracy consistently across both architectures, and
improves energy accuracy in most cases, with the exception of MACE's energy RMSE. This suggests
that even light-touch curation, removing well under 1\% of training structures, provides a measurable
benefit, particularly for force prediction, with a slight additional gain in training speed due to the smaller dataset size.

% Fig.~\ref{fig:curated-vs-uncurated-curve-force-mae}

\begin{table}[h!]
  \centering
\caption{Validation force errors (meV/\AA) for MACE and ViSNet trained on curated vs.\ uncurated SPICE2, computed at each run's best epoch. Reported values are averaged over two independent initialization seeds per model/dataset combination.}
  \label{tab:force-curated-vs-uncurated-mlflow}
  \vspace{0.3cm}
\renewcommand{\arraystretch}{1.3}
\setlength{\tabcolsep}{6pt}
  \begin{tabular}{lcccc}
    \toprule
    & \multicolumn{2}{c}{MAE} & \multicolumn{2}{c}{RMSE} \\
    \cmidrule(lr){2-3} \cmidrule(lr){4-5}
    Model & Curated & Uncurated & Curated & Uncurated \\
    \midrule
    MACE & 17.2 & 17.9 & 34.4 & 36.0 \\
    ViSNet & 16.5 & 17.3 & 31.5 & 32.6 \\
    \bottomrule
  \end{tabular}
\end{table}

\begin{table}[h!]
  \centering
\caption{Validation energy errors (meV) for MACE and ViSNet trained on curated vs.\ uncurated SPICE2, computed at each run's best epoch. Reported values are averaged over two independent initialization seeds per model/dataset combination.}
  \label{tab:energy-curated-vs-uncurated-mlflow}
  \vspace{0.3cm}
\renewcommand{\arraystretch}{1.3}
\setlength{\tabcolsep}{6pt}
  \begin{tabular}{lcccc}
    \toprule
    & \multicolumn{2}{c}{MAE} & \multicolumn{2}{c}{RMSE} \\
    \cmidrule(lr){2-3} \cmidrule(lr){4-5}
    Model & Curated & Uncurated & Curated & Uncurated \\
    \midrule
    MACE & 33.0 & 33.1 & 62.0 & 59.7 \\
    ViSNet & 24.4 & 26.9 & 46.2 & 51.0 \\
    \bottomrule
  \end{tabular}
\end{table}

\section{Conclusion and Outlook}

In summary, MLIP performance is fundamentally constrained not by how much data is available, but by how well that data spans the relevant high-dimensional potential energy surface, especially in regions that equilibrium MD rarely visits. Addressing this challenge requires infrastructure that makes large, heterogeneous chemical datasets easy to access, curate, and reshape around the specific needs of a given modeling task, rather than around the constraints of any single source dataset's original format.
In this work, we introduced \textit{ChemReporter}, a modular, method-agnostic framework that converts raw chemical and materials datasets into a unified, queryable Parquet-based representation, and provides a decoupled pipeline for processing, querying, and exporting MLIP-ready training data.

Looking ahead, we plan to expand \textit{ChemReporter}'s built-in library of sampling and filtering algorithms, as well as the range of precomputed properties available for querying, further broadening the tool's utility for constructing well-curated, task-specific training sets.

\bibliographystyle{IEEEtranN}
\bibliography{references}

\newpage

%%%%%%%%%%%%%%%%%%%%%%%%%%%%%%%%%%%%%%%%%%%%%%%%%%%%%%%%%%%%

\newpage
\appendix

\section{MACE Training Details} \label{app:mace_training}

We trained a MACE-based MLIP as implemented in the \textit{mlip} library~\cite{brunken2026machine}. The model comprises two message-passing layers with 128 uncoupled feature channels, spherical harmonics up to $l_{\max}=3$, and node features restricted to the same maximum degree ($L=3$). The message function uses a maximum correlation order of $\nu=2$, corresponding to 3-body message expansions. Radial features are represented using 16 radial basis functions with a polynomial envelope.

The model was trained on a subset of the OMol25 dataset (described in section~\ref{sec:exp_setup} of the main text) for 220 epochs with a batch size of 128. The training objective minimized the MSE of total energies and atomic forces using a dynamic, epoch-dependent "weight-flipping" schedule: during the first 115 epochs, the force penalty was heavily weighted (1000.0 vs.\ 40.0 for energy); for the final 105 epochs, the loss was flipped to prioritize energy fitting.

% \section{ViSNet Training Details} \label{app:visnet_training}

% We trained the MLIP model using the ViSNet architecture as implemented in the \texttt{mlip} package~\cite{brunken2026machine}. The model comprises 4 message-passing layers with 128 feature channels and 8 attention heads, employing spherical harmonics up to a maximum degree of $l_{\max} = 2$. Radial environments are represented using 32 radial basis functions. Sigmoid Linear Unit (\textsc{SiLU}) activation functions are applied uniformly across the embedding, attention, and output blocks. No trainable radial basis functions or vector-norm reweightings were used.

% The model was trained over a total of 220 epochs with a batch size of 128. The training objective minimized the mean squared error (\textsc{MSE}) of both total energies and atomic forces, following an epoch-dependent weight-scheduling strategy identical to that used for MACE:
% The force penalty was weighted heavily relative to the energy loss ($w_{\text{force}} = 1000.0$ vs.\ $w_{\text{energy}} = 40.0$) to prioritize the fit of local atomic environments.
% The weighting schedule was inverted to prioritize precise global energy predictions in epochs 115 to 220.

\section{ViSNet Training Details}
\label{app:visnet_training}

We trained a ViSNet-based MLIP as implemented in the \textit{mlip} library~\cite{brunken2026machine}. The model comprises 4 message-passing layers with 128 feature channels and 8 attention heads, employing spherical harmonics up to a maximum degree of $l_{\max} = 2$. Radial environments are represented using 32 radial basis functions.

The model was trained on a subset of the OMol25 dataset (described in section~\ref{sec:exp_setup} of the main text) for 220 epochs with a batch size of 128. The training objective minimized the MSE of total energies and atomic forces using the same dynamic, epoch-dependent "weight-flipping" schedule as MACE (see Appendix~\ref{app:mace_training}).

\section{Training Curves} 
\label{app:training_curves}

\begin{figure}[h!]
  \centering
  \includegraphics[width=0.85\linewidth]{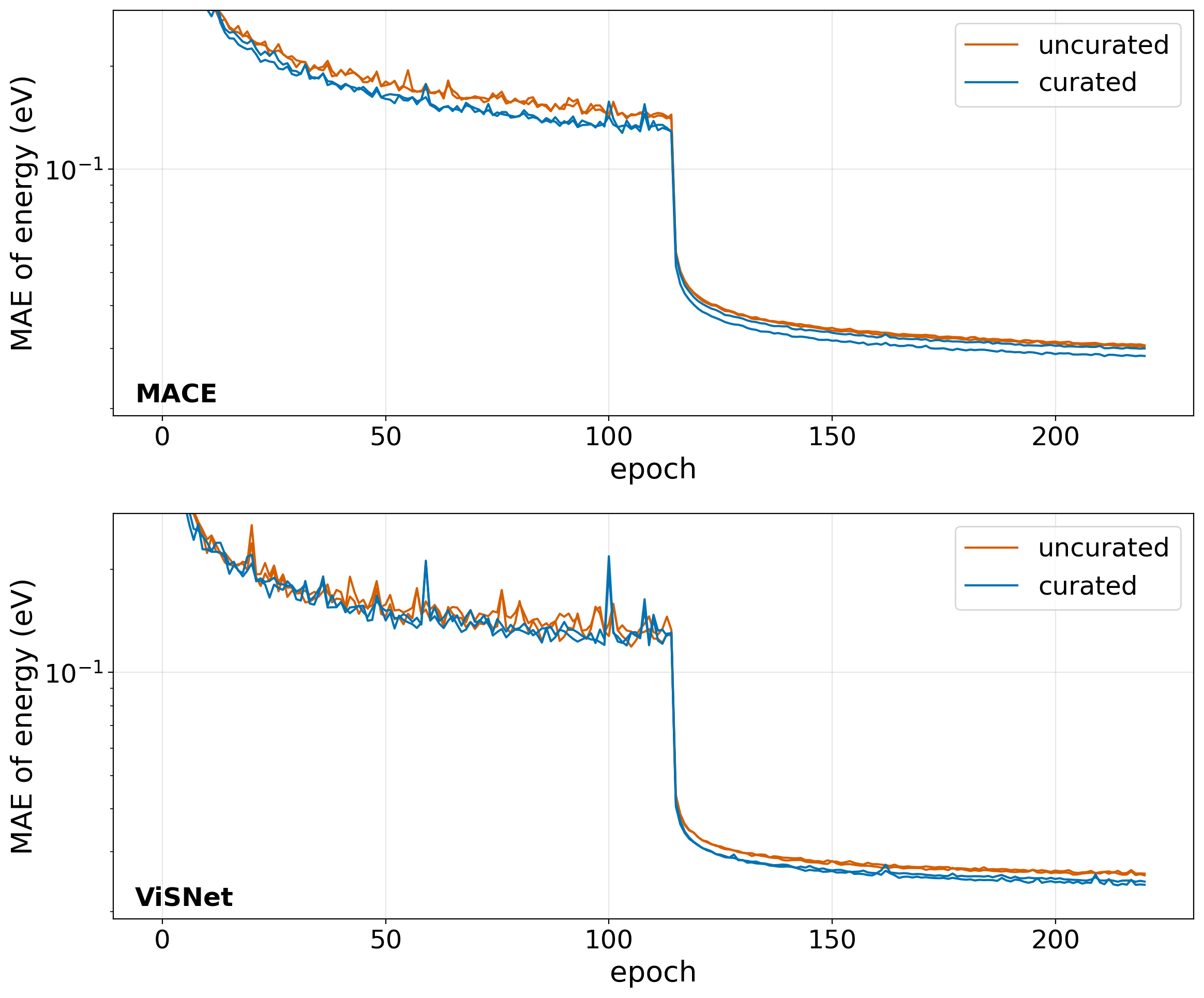}
\caption{Validation energy MAE as a function of training epoch for MACE (top) and ViSNet (bottom), comparing models trained on curated vs.\ uncurated SPICE2. Across both architectures, models trained on the curated dataset achieve consistently lower validation energy error throughout training. The visible discontinuity around epoch 115 corresponds to the loss weight-flipping schedule. Two lines per dataset correspond to results from two independent initialization seeds.}
  \label{fig:curated-vs-uncurated-curve-energy-mae}
\end{figure}

\begin{figure}[b]
  \centering
  \includegraphics[width=0.85\linewidth]{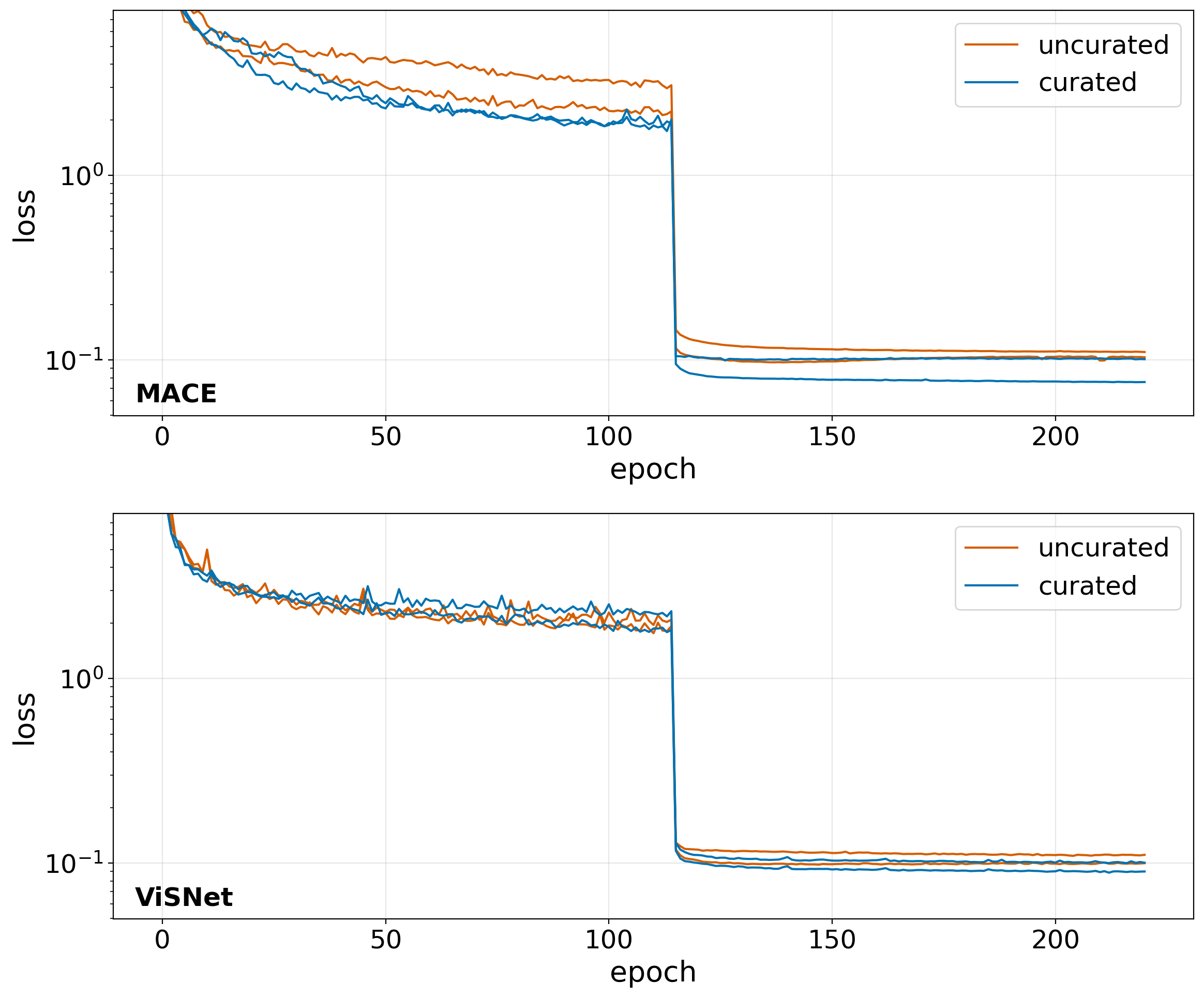}
\caption{Validation loss as a function of training epoch for MACE (top) and ViSNet (bottom), comparing models trained on curated vs.\ uncurated SPICE2. The visible discontinuity around epoch 115 corresponds to the loss weight-flipping schedule. Two lines per dataset correspond to results from two independent initialization seeds.}
  \label{fig:curated-vs-uncurated-curve-loss}
\end{figure}

In this section, we present additional training curves with further metrics, i.e., energy MAE (see Figure~\ref{fig:curated-vs-uncurated-curve-energy-mae}) and loss (see Figure~\ref{fig:curated-vs-uncurated-curve-loss}).

% Fig.~\ref{fig:curated-vs-uncurated-curve-energy-mae}
% Fig.~\ref{fig:curated-vs-uncurated-curve-energy-mae-per-atom}

% To account for stochasticity in weight initialization and mini-batch sampling, each training configuration was executed independently across two different random seeds.

%%%%%%%%%%%%%%%%%%%%%%%%%%%%%%%%%%%%%%%%%%%%%%%%%%%%%%%%%%%%

% \clearpage
% \newpage
% \input{checklist.tex}

\end{document}